\documentclass[prd,showpacs,preprint]{revtex4}
\usepackage{amssymb}
\usepackage{amsmath}
\usepackage{graphicx}
\usepackage{multirow}

\begin{document}

\title{Signature of Large Extra Dimensions from Z boson pair production at the CERN Large Hadron Collider}
\author{Jun Gao}
\author{Chong Sheng Li}
\email{csli@pku.edu.cn}
\author{Xiangdong Gao}
\author{Jia Jun Zhang}
\affiliation{Department of Physics and State Key Laboratory of
Nuclear Physics and Technology, Peking University, Beijing 100871,
China}

\date{\today}


\begin{abstract}
We study the Z boson pair production mediated by the Kaluza-Klein
(KK) graviton in large extra dimensions (LED) at the CERN Large
Hadron Collider (LHC). We use the partial wave unitarity to discuss
the constraints on the process energy scale in order to give a
self-consistent calculation. We find that the LED contributions can
enhance the Z boson pair production cross sections significantly
when the fundamental scale $M_S$ of the large extra dimensions is up
to several TeV. We also show that the kinematic distributions of the
LED signals are greatly different from the standard model ones and
the LHC can probe the $M_S$ values up to $4.3\sim 5.6$ TeV at
$3\sigma$ level depending on the number of the extra dimensions.
\end{abstract}

\maketitle

\section{introduction}
The idea that quantum gravity can appear at the TeV energy scale
well below the Planck mass $M_{\rm P}\sim 1.2\times 10^{19}$GeV was
proposed in the
1990's~\cite{Antoniadis:1990ew,ArkaniHamed:1998rs,Randall:1999ee}.
The large extra dimensions(LED) model~\cite{ArkaniHamed:1998rs}
introduced by Arkani-Hamed, Dimopoulos and Davli has attracted much
attention as the presence of large extra dimensions brings a new
solution to the hierarchy problem. In the LED model, $n$ extra
spatial dimensions are compactified on a torus with common
circumference $R$, and a 3-brane is introduced in which the standard
model (SM) particles live. The SM gauge interactions are confined to
this brane and only the gravity can propagate in the extra
dimensions. Then the four-dimensional Planck scale $M_{\rm{P}}$ is
no longer the relevant scale but is related to the fundamental scale
$M_S$ as follows~\cite{Csaki:2004ay}:
\begin{eqnarray}\label{scale}
M^2_{\rm{P}}\sim M^{n+2}_SR^n,
\end{eqnarray}
where $M_S\sim {\rm TeV}$. According to Eq.~(\ref{scale}),
deviations from the usual Newtonian gravitational force law can be
expected at distances smaller than $R\sim
2\times10^{-17+32/n}\rm{cm}$~\cite{Csaki:2004ay}. For $n\geq 2$, LED
is consistent with the current experiments since gravitational
forces currently are only well probed at distances about 40 $\mu
m$~\cite{Kapner:2006si} (However for $n=2$, there are constraints
arising from, e.g., supernova cooling~\cite{Hanhart:2001fx}, which
require $M_S\gtrsim 14 \rm{TeV}$ if $n=2$). Recently, there are also
some new constraints $M_S\gtrsim {\rm 1TeV}$~\cite{Krutelyov:2008nb}
from direct search at the Tevatron.

The gravitational couplings of the SM fields are given by
\begin{equation}
\int d x^4\sqrt{-\hat g}{\mathcal L}_{SM}(\hat g,S,V,F),
\end{equation}
where $\hat g$ is the induced metric in four dimensions. Expanding
$\hat g$ with the gravitational fields, we obtain
\begin{equation}\label{cou}
-\frac{\kappa}{2}\sum_{\vec{k}}\int d x^4
(\tilde{h}^{\mu\nu,\vec{k}}T_{\mu\nu}+\omega
\tilde{\phi}^{\vec{k}}T^{\mu}_{\ \nu}),
\end{equation}
where $\kappa=\sqrt{16\pi}/M_P$, $\omega=\sqrt{2/(3n+6)}$,
$T_{\mu\nu}$ is the energy-momentum tensor and
$\tilde{h}^{\mu\nu,\vec{k}}$, $\tilde{\phi}^{\vec{k}}$ are the four
dimensional Kaluza-Klein (KK) gravitons and KK scalars,
respectively. From Eq.~(\ref{cou}) it can be seen that the
individual KK mode couples with a strength $1/M_P$ to the SM fields.
However, since there are many KK modes, the total coupling strength
is of the order $1/M_S$ after summing up all of them. For the
phenomenology studies, we use the convention introduced in
Ref.~\cite{Han:1998sg}, where $M_S$ is chosen to satisfy
\begin{equation}
\kappa^2 R^n=8\pi(4\pi)^{n/2}\Gamma(n/2)M_S^{-(n+2)}.
\end{equation}

In this year, the LHC will start running and begin searching for the
Higgs boson, further testing the SM, and also searching for new
physics beyond the standard model. The $Z$ boson pair production is
an important process at the LHC and has been studied extensively. In
the SM the $Z$ boson pair can be produced directly through
electroweak interactions with a total cross section of about 17
pb~\cite{Campbell:1999ah} at the LHC, corresponding to
$1.7\times10^6$ events with an integrated luminosity of 100 $\rm
fb^{-1}$. With such a high production rate, it can serve as a probe
for testing SM at TeV scale and also be an important background for
new physics signals. Besides, the $Z$ boson pair production can get
additional contributions from new physics, for example, from the
tree level processes in extra dimension
models~\cite{Atwood:1999cy,Park:2001vk}, little Higgs
model~\cite{Hill:2007zv} and from loop level process in
supersymmetry model~\cite{Berger:1998vx}, thus it can be used as a
probe of new physics beyond standard model.

Usually, the most promising processes to test the LED model at the
LHC are dilepton production, diphoton production, dijet production
and single vector boson or jet production associated with KK
graviton~\cite{Giudice:1998ck,Atwood:1999cy}. However, the Z boson
pair production also can be an important process especially for low
$M_S$ value, the total cross sections and invariant mass
distributions of this process have been given in
Ref.~\cite{Atwood:1999cy}. In this paper we will further study the Z
boson pair production through the s-channel KK graviton exchange at
the LHC. We focus on the 4 charged lepton final states and perform a
Monte Carlo simulation to see the new physics effects on several
kinematic distributions. Finally, we also show the LHC discovery
potential of $M_S$ with different values of $n$.

The arrangement of this paper is as follows. In Sec.~II, we give the
helicity amplitudes and discuss the unitarity constraints on the
process energy scale. Sec.~III shows the results of the total cross
sections and Sec.~IV contains the simulation results of several
kinematic distributions and also the LHC reach of $M_S$. Sec.~V is a
brief conclusion.

\section{helicity amplitudes and unitarity constraints}
In the LED model the Z boson pair can be produced through KK
graviton exchange from the $gg$ fusion and $q\bar q$ annihilation,
respectively, at the LHC. In order to calculate the relevant
amplitudes we need to sum over all the KK graviton propagators,
\begin{equation}
D(s)=\sum_{\vec k} \frac{i}{s-m_{\vec k}^2+i\epsilon}.
\end{equation}
Since the KK graviton mass separation of $\mathcal{O}(1/R)$ is much
smaller than all the other physical scales involved, we can obtain
the above sum in the continuum limit~\cite{Han:1998sg},
\begin{equation}
D(s)=\frac{s^{n/2-1}}{\Gamma(n/2)}\frac{R^n}{(4\pi)
^{n/2}}[\pi+2i{\rm I}(\Lambda/\sqrt{s})],
\end{equation}
where $\Lambda\sim\mathcal{O}(M_S)$ is the explicit ultraviolet
cutoff scale of the effective field theory, and
\begin{eqnarray}\label{ifactor}
{\rm I}(\Lambda/\sqrt{s})&=&-\sum^{n/2-1}_{k=1}\frac{1}{2k}\left
(\frac{\Lambda}{\sqrt{s}}\right)^{2k}-\frac{1}{2}\log{\left
(\frac{\Lambda^2}{s}-1\right)}\quad (\rm{n=even}) \nonumber \\
&=&-\sum_{k=1}^{(n-1)/2}\frac{1}{2k-1}\left(\frac{\Lambda}{\sqrt{s}}
\right)^{2k-1}+\frac{1}{2}\log{\left(\frac{\Lambda+\sqrt{s}}
{\Lambda-\sqrt{s}}\right)}\quad (\rm{n=odd}).
\end{eqnarray}
At the LHC the subprocess energy is comparable to $M_S$, so we need
to keep all the terms in Eq.~(\ref{ifactor}).

\begin{figure}[h]
\includegraphics[width=0.9\textwidth]{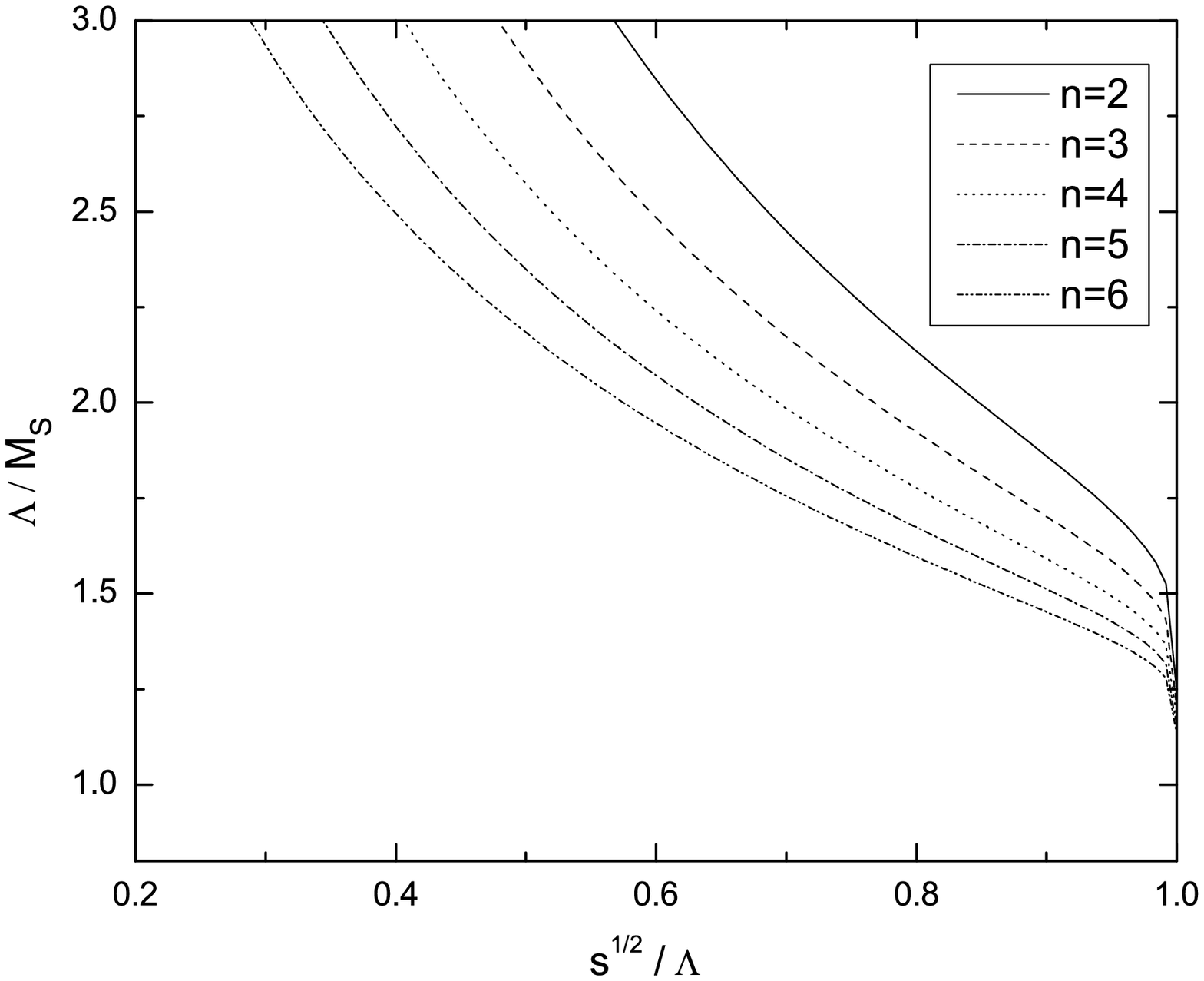}
\caption[]{Partial wave unitarity constraints on the ratios $\sqrt
s/M_S$ and $\Lambda/M_S$, the allowed region is to the left of and
below the curves.} \label{f1}
\end{figure}

Since the scattering amplitudes grow quickly with $\sqrt s$, we
should cut off the subprocess energy at some scale below $\Lambda$
in order not to violate the unitarity. And we use the partial wave
unitarity to discuss the constraint on $\sqrt s$. The J-partial wave
amplitudes are given by
\begin{equation}
a_{\mu\mu'}^J=\frac{1}{64\pi}\int_{-1}^1 d\cos \theta \
d_{\mu\mu'}^J(\cos \theta)\
\left[\mathcal{M}^{\lambda_1\lambda_2\lambda_3\lambda_4}\right],
\end{equation}
where $\mu=\lambda_1-\lambda_2,\mu'=\lambda_3-\lambda_4$, $\theta$
is the scattering angle and $d^J_{\mu\mu'}$ is the Wigner
function~\cite{Amsler:2008zzb}. For the subprocess $gg\to ZZ$, the
individual non-vanishing helicity amplitudes are
\begin{eqnarray}
&i{\mathcal M}_{+-+0}=&-i{\mathcal M}_{+-0-}=2 \sqrt{2}i\delta_{cd}
{\rm M_S}^{-n-2} {\rm m_z} \pi  s^{\frac{n+1}{2}} (1+\cos (\theta ))
 \sin (\theta ){\mathcal A};\nonumber\\
&i{\mathcal M}_{+--0}=&-i{\mathcal M}_{+-0+}=2 \sqrt{2}i\delta_{cd}
{\rm M_S}^{-n-2} {\rm m_z} \pi  s^{\frac{n+1}{2}} (1-\cos (\theta ))
 \sin (\theta ){\mathcal A};\nonumber\\
&i{\mathcal M}_{+-++}=&i{\mathcal M}_{+---}=4i\delta_{cd} {\rm
M_S}^{-n-2} {\rm m_z}^2 \pi s^{n/2}
\sin ^2(\theta ){\mathcal A};\nonumber\\
&i{\mathcal M}_{+-+-}=&4 i\delta_{cd}{\rm M_S}^{-n-2} \pi
s^{\frac{n}{2}+1}\cos ^4\left(\theta
/2\right){\mathcal A};\nonumber\\
&i{\mathcal M}_{+--+}=&4 i\delta_{cd}{\rm M_S}^{-n-2} \pi
s^{\frac{n}{2}+1} \sin ^4\left(\theta
/2\right){\mathcal A};\nonumber\\
&i{\mathcal M}_{+-00}=&-i\delta_{cd}{\rm M_S}^{-n-2} \pi  s^{n/2}
\left(4 {\rm m_z}^2+s\right)\sin ^2(\theta ){\mathcal A},
\end{eqnarray}
where $c,d$ are the gluon color indices and
\begin{equation}
{\mathcal A}=-2{\rm I}(\Lambda/\sqrt s)+i\pi.
\end{equation}
Since the energy scale we considered here is much larger than $m_Z$,
so we can neglect all the terms that are proportional to $m_Z$. Then
the main contributions arise from ${\mathcal M}_{+-+-}$, ${\mathcal
M}_{+--+}$ and ${\mathcal M}_{+-00}$, and the only non-vanishing
partial wave amplitudes correspond to the $J=2$ partial wave, which
are given by
\begin{equation}
a^2_{2\pm 2}=\frac{1}{40} \left(\frac{\sqrt s}{M_S}\right)^{n+2}
{\mathcal A}, \ \ a^2_{20}=-a^2_{22}/ \sqrt 6.
\end{equation}
All the above amplitudes $a^2_{22}$, $a^2_{2-2}$ and $a^2_{20}$
contribute to the imaginary part of $gg$ elastic scattering
amplitudes according to the optical theorem, so the partial wave
unitarity leads to
\begin{equation}\label{uni}
|a^2_{22}|^2+|a^2_{2-2}|^2+|a^2_{20}|^2<1.
\end{equation}
The partial wave amplitudes of $q\bar q \to ZZ$ lead to weaker
constraints compared with the ones of $gg\to ZZ$, and are thus not
shown here. In Fig.~\ref{f1} we show the unitarity constraints on
$\sqrt s$ with different $\Lambda/M_S$ values according to
Eq.~(\ref{uni}). We can see that the constraints on $\sqrt s$ become
strong for large $\Lambda/M_S$ values. In the following calculations
we set $\Lambda=M_S$ and choose the energy cut-off to be $0.9M_S$
for simplicity, which is allowed by the unitarity constraints.

\section{cross section calculations}
The additional contributions to the partonic differential cross
sections of the Z boson pair production in the LED model are given
by
\begin{eqnarray}
\frac{d\hat{\sigma}_{q\bar q}}{d\cos
\theta}&=&\frac{\hat{s}\sqrt{1-4x}\ a^2(1+b^2){\mathcal
F}(n){\mathcal
T}(n)}{24}\nonumber\\
&&\times\left\{\frac{z^4\left(1-4x\right)^2+10z^2x\left(1-4x\right)
-2x\left(1-4x\right)-1}
{\left(1-z^2\right)\left(1-4x\right)+4x^2}\right\}\nonumber\\
&&+\frac{\pi \hat{s}^3\sqrt{1-4x}
\mathcal{F}^2(n)\left(\pi^2+4\mathcal{T}^2(n)\right)}{384}\nonumber\\
&&\times\left\{-3z^4\left(1-4x\right)^2+z^2\left(1-4x\right)
\left(1-12x\right)+2(1+4x) \right\},
\end{eqnarray}
and
\begin{eqnarray}
\frac{d\hat{\sigma}_{gg}}{d\cos \theta}&=&\frac{
\pi\hat{s}^3\sqrt{1-4x}
\mathcal{F}^2(n)\left(\pi^2+4\mathcal{T}^2(n)\right)}{1024}\nonumber\\
&&\times\left\{
3z^4\left(1-4x\right)^2+2z^2\left(1-4x\right)\left(5+12x\right)
+\left(1+12x\right)\left(3+4x\right)\right\},
\end{eqnarray}
with
\begin{equation}
{\mathcal F}(n)=\hat {s}^{n/2-1}/M_S^{n+2}, \ {\mathcal
T}(n)=I(M_S/\sqrt{\hat s}),
\end{equation}
where $x=m_Z^2/\hat{s}$, $z=\cos\theta$, $a=e/(4\sin \theta_W \cos
\theta_W)$, and $b=-1+8\sin^2 \theta_W/3$ or $-1+4\sin^2
\theta_W/3$, for up-type and down-type quark, respectively. Note
that the factors ${\mathcal F}(n)$ and ${\mathcal T}(n)$ satisfy
\begin{equation}
\kappa^2 D(\hat s)=8\pi{\mathcal F}(n)(\pi+2i{\mathcal T}(n)).
\end{equation}

\begin{figure}[h!]
\includegraphics[width=0.9\textwidth]{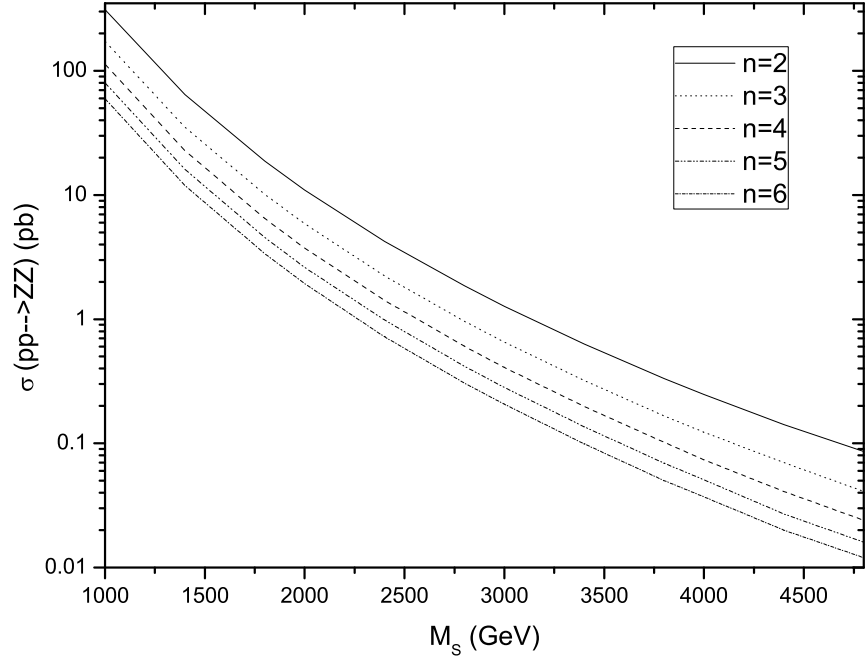}
\caption[]{The total cross sections of the Z boson pair production
from LED contributions as functions of $M_S$ at the LHC.} \label{f2}
\end{figure}

In the numerical calculations we use the CTEQ6Ll PDF
set~\cite{Pumplin:2002vw} and choose the factorization scale to be
the Z boson mass. To estimate the uncertainty of the tree-level
calculations, we vary the factorization scale from $m_Z$ to 1 TeV,
and find that the uncertainty of the total cross section is about
$\lesssim50\%$, corresponding to a rather small shift of $M_S$.
Fig.~\ref{f2} shows the total cross sections in the LED model as
functions of $M_S$. We can see that the cross sections drop quickly
with the increasing $M_S$, but can still reach hundreds of $\rm fb$
for $M_S=4 {\rm TeV}$. At low $ M_S$ values, the $gg$ fusion channel
is dominant, while for large $ M_S$ the $q\bar q$ annihilation
contributions become important and are comparable to the ones of
$gg$ fusion channel.
\section{Monte Carlo Simulations}
We only consider the 4 charged leptons final state since at such
high energy scale the two jets from highly boosted Z boson decay are
hard to distinguish and the jet final states will suffer from large
backgrounds also. The partonic level events are generated with
COMPHEP 4.4~\cite{Boos:2004kh} and PYTHIA6.4~\cite{Sjostrand:2006za}
is used to treat initial and final state radiations. We smear the
energy and direction of the charged leptons by a Gaussian
distribution with the errors given by~\cite{Bayatian:2006zz}
\begin{eqnarray}
&&\Delta E_l/E_l=0.1/\sqrt{E_l/{\rm GeV}}\oplus0.007,\nonumber\\
&&\Delta \eta_l/\eta_l=\Delta \phi_l/\phi_l=0.001.
\end{eqnarray}
Here we do not distinguish between electrons and muons for
simplicity. We impose the basic acceptance cuts as
follows~\cite{Bayatian:2006zz}:
\begin{equation}
p_T^l>15{\rm GeV}, |\eta_l|<2.4,\ {\rm and}\ \Delta R_{ll}>0.1.
\end{equation}
For the reconstruction of the two $Z$ bosons we require the lepton
combination of minimizing the following quantity
\begin{equation}
\Delta \equiv \sqrt{(m_{l_1l_2}-m_Z)^2+(m_{l_3l_4}-m_Z)^2}.
\end{equation}

\begin{figure}[h!]
\includegraphics[width=0.9\textwidth]{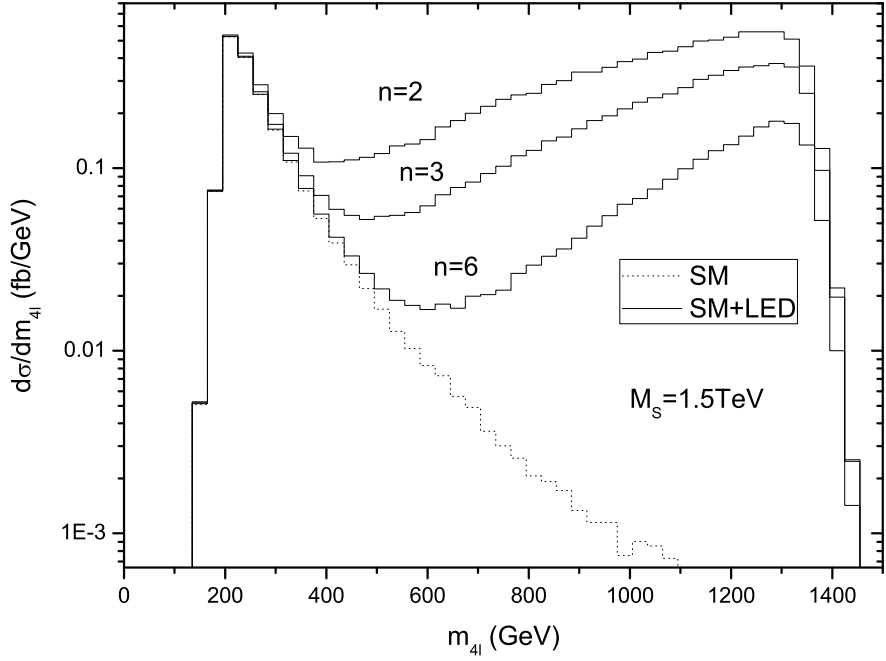}
\includegraphics[width=0.9\textwidth]{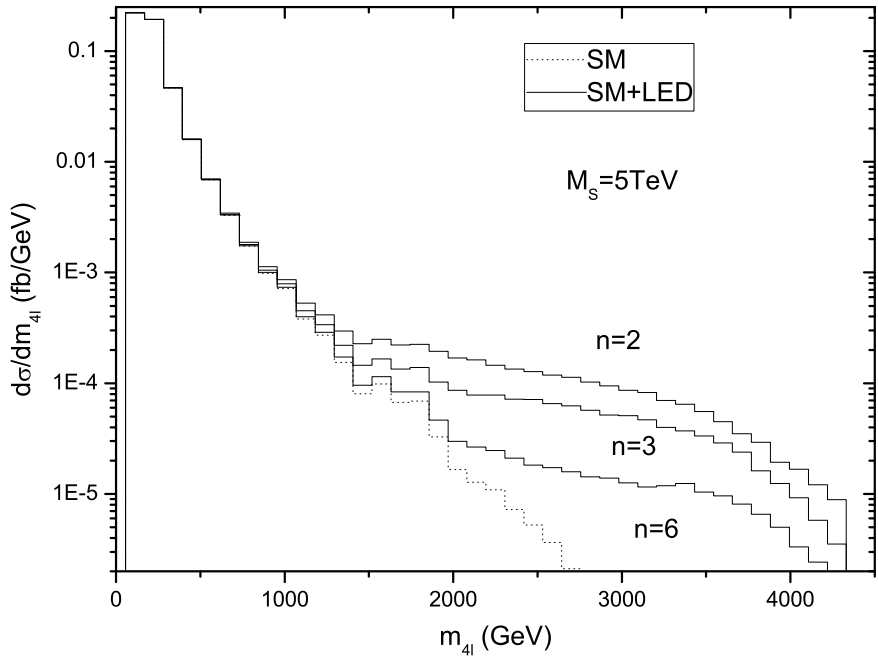}
\caption[]{The invariant mass distributions of the 4 leptons.}
\label{f3}
\end{figure}

\begin{figure}[h!]
\includegraphics[width=0.9\textwidth]{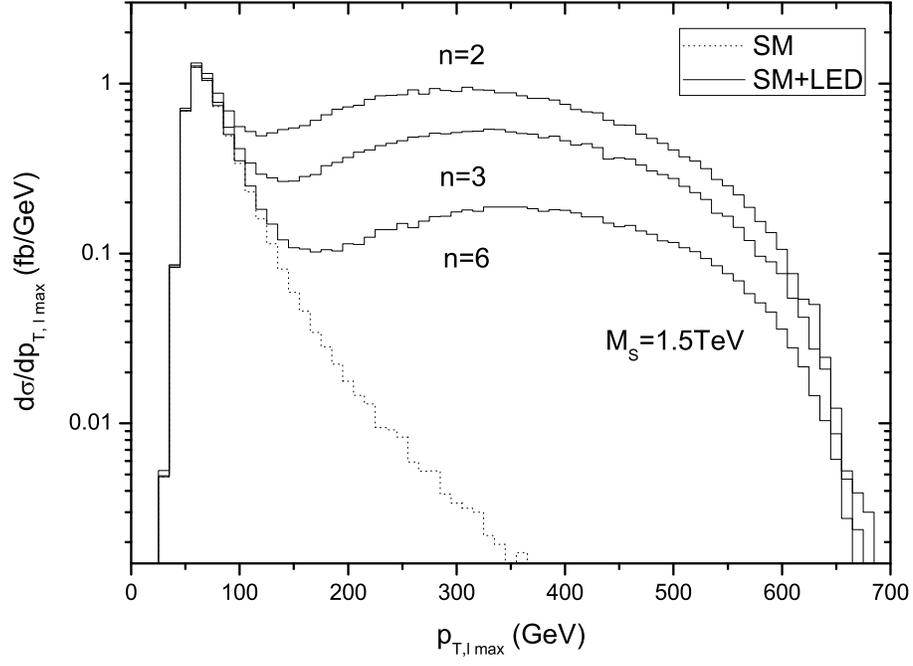}
\includegraphics[width=0.9\textwidth]{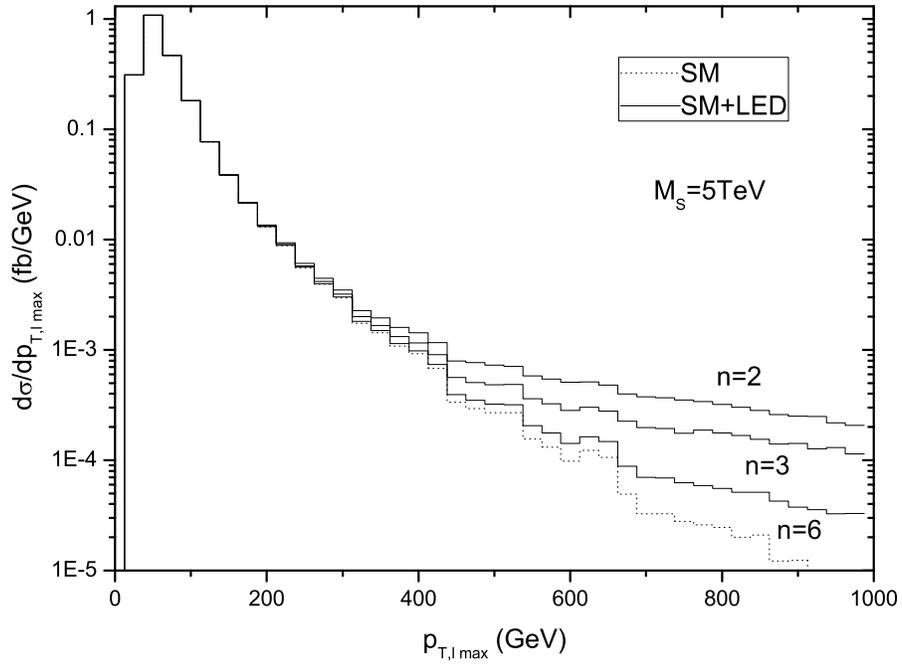}
\caption[]{The transverse momentum distributions of the leading
lepton.} \label{f4}
\end{figure}

\begin{figure}[h!]
\includegraphics[width=0.9\textwidth]{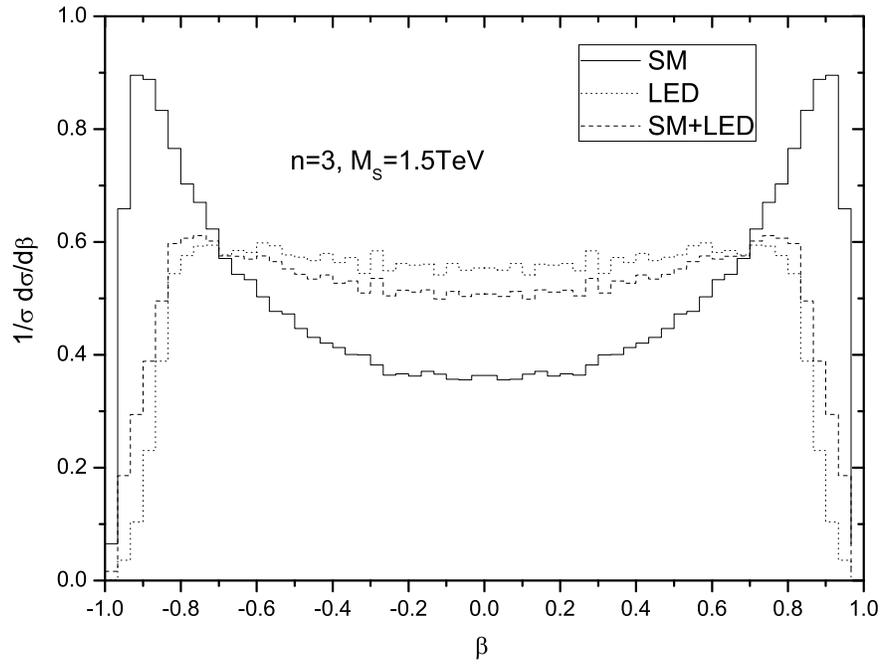}
\caption[]{The normalized velocity distributions of the rest frame
of the Z pair.} \label{f5}
\end{figure}

\begin{figure}[h!]
\includegraphics[width=0.9\textwidth]{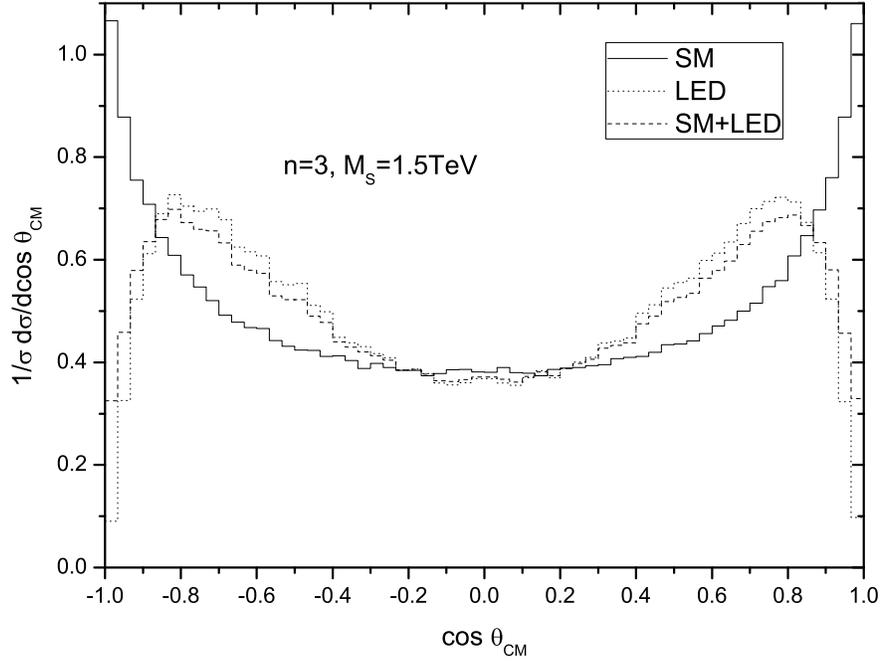}
\caption[]{The normalized polar angle distributions of the
reconstructed Z boson in the rest frame of the Z pair.} \label{f6}
\end{figure}

In Figs.~\ref{f3} and ~\ref{f4}, we show the 4 lepton invariant mass
distributions and the leading lepton $p_T$ distributions,
respectively. It can be seen that the LED signals prefer the high
invariant mass and high $p_T$ regions. And for $M_S=1.5{\rm TeV}$,
the LED contributions can change the shape of the distributions
significantly, while for $M_S=5{\rm TeV}$, the changes of the
distributions are relatively small, leading to long tails in the
distributions. Fig.~\ref{f5} gives the velocity or boost factor
$\beta$ distributions of the rest frame of the Z pair. We can see
that the SM distribution will have two peaks in the forward and
backward region since the Z boson pair can only be produced through
$q\bar q$ annihilation in the SM. But for the LED contributions,
$gg$ fusion is dominant at low $M_S$ values, thus the distribution
will be flat and drop in both ends. In Fig.~\ref{f6} we display the
polar angle distributions of the reconstructed Z boson in the rest
frame of the Z pair. Compared with the SM case, the LED signal falls
down in both ends since the leptons from the decay of highly boosted
Z boson in these regions fail the rapidity cut.

\begin{figure}[h!]
\includegraphics[width=0.9\textwidth]{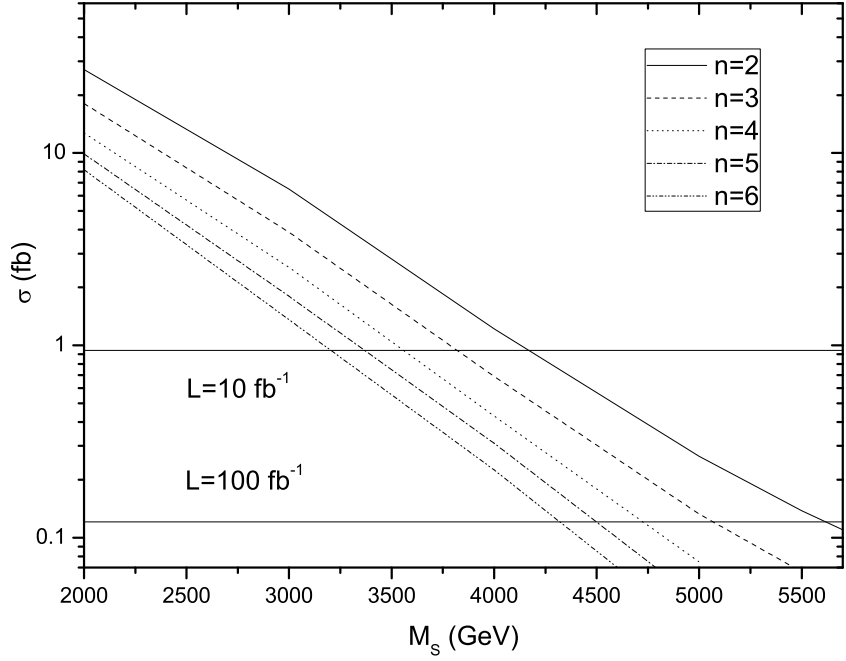}
\caption[]{The total cross sections of the LED signals after all the
cuts as functions of $M_S$. The horizontal lines indicate the cross
sections needed for a $3\sigma$ detection of the signal.} \label{f7}
\end{figure}

In order to further investigate the LHC reach of $M_S$, we introduce
an invariant mass cut $m_{4l}>1.5{\rm TeV}$ and use a simple event
counting method to calculate the signal significance. The Z boson
pair production cross section in the SM is normalized to $17 {\rm
pb}$ based on the next-to-leading order result of
Ref.~\cite{Campbell:1999ah}. Table~\ref{t1} gives some results of
the cut efficiencies, signal significance and total cross sections
after all the cuts. It can be seen that the invariant mass cut
reduces the SM production rate significantly while only has small
influence on the LED signals, especially for large $M_S$ values. In
Fig.~\ref{f7} we show the LED signal cross sections as functions of
$M_S$ after all the cuts, and also the LHC reach of $M_S$. Generally
speaking, for the Z boson pair production process, the LHC can probe
the $M_S$ values up to $4.3\sim 5.6$ TeV for ${\rm n}=6\sim 2$ at
$3\sigma$ level, assuming an integrated luminosity of 100 $fb^{-1}$.

It should be noted that since we cut off the subprocess energy at
some scale, we didn't include the LED contributions beyond the
cutoff in our calculations, because they depend on the unknown
ultraviolet completion of the LED theory. For low $M_S$ values, like
$M_S=1.5{\rm TeV}$, these contributions may be comparable or even
larger than the ones we have considered, which may change the total
cross sections and all the distributions. On the other hand, for
high $M_S$ values, like $M_S=5{\rm TeV}$, the LED contributions
beyond the cutoff are small due to the suppression of the parton
distribution functions and the limited energy of the LHC, and can be
neglected. So these contributions will not change the above results
of LHC reach for $M_S$, which is about 5 TeV.

\begin{table*}
\begin{center}
\begin{tabular}{|c|c|c|c|c|c|c|c|c|c|c|}
\hline  \multirow{2}{*}{cut efficiency}&\multicolumn{3}{|c|}{$\ \rm
M_S=2TeV$}& \multicolumn{3}{|c|}{$\ \rm
M_S=3TeV$}&\multicolumn{3}{|c|}{$\ \rm
M_S=4TeV$}&\multirow{2}{*}{SM}
\\
\cline{2-10}     & n=2& n=3& n=6& n=2& n=3& n=6&n=2& n=3& n=6&
\\
\hline basic cuts &0.73&0.73&0.76&0.76&0.78&0.79&0.72&0.74&0.75&0.35
\\
\hline $m_{4l}>1.5{\rm TeV}$
&0.25&0.30&0.42&0.51&0.59&0.66&0.49&0.56&0.61&0.00027
\\
\hline total cross section (fb)   &27.2&18.1&
8.2&6.49&3.87&1.36&1.22&0.69&0.22&0.046
\\
\hline $S/\sqrt{S+B}$        &52&42&28&25&20&12&11&8.1&4.3&
\\
\hline
\end{tabular}
\end{center}
\caption{Cut efficiencies, signal significance and total cross
sections after all the cuts with different parameters, assuming an
integrated luminosity of 100 $fb^{-1}$.} \label{t1}
\end{table*}

\section{conclusions}

In conclusion, we have studied the Z boson pair production mediated
by the KK graviton in large extra dimensions at the LHC. Since the
scattering amplitudes grow quickly with the energy, we used the
partial wave unitarity to discuss the constraints on the process
energy scale. We found that the LED contributions can enhance the Z
boson pair production cross sections significantly when the
fundamental scale $M_S$ of the large extra dimensions is up to
several TeV, and the kinematic distributions of the LED signals are
greatly different from the SM ones through Monte Carlo simulations.
Finally, we investigated the LHC reach of $M_S$ and found that the
LHC can probe the $M_S$ values up to $4.3\sim 5.6$ TeV for ${\rm
n}=6\sim 2$ at $3\sigma$ level assuming an integrated luminosity of
100 $fb^{-1}$.

\begin{acknowledgments}
We would like to thank C.-P.Yuan for useful discussions. This work
was supported in part by the National Natural Science Foundation of
China, under Grants No.10721063 and No.10635030.
\end{acknowledgments}

\newpage
\bibliography{ledzz}
\newpage

\end{document}